# Application Specific Cache Simulation Analysis for Application Specific Instruction set Processor

Ravi Khatwal
Research scholar
Department Of Computer science,
Mohan LaL Sukhadia University,
Udaipur, India.

Manoj Kumar Jain, Ph. D
Associate Professor
Department Of Computer science,
Mohan LaL Sukhadia University,
Udaipur, India.

## ABSTRACT
An Efficient Simulation of application specific instruction-set processors (ASIP) is a challenging onus in the area of VLSI design. This paper reconnoiters the possibility of use of ASIP simulators for ASIP Simulation. This proposed study allow as the simulation of the cache memory design with various ASIP simulators like **Simple scalar** and **VEX.** In this paper we have implemented the memory configuration according to desire application. These simulators performs the cache related results such as cache name, sets, cache associativity, cache block size, cache replacement policy according to specific application.

## 1. INTRODUCTION
ASIPs are the challenging task in the area of high performance embedded system design. ASIP performs the target architecture such big-endian and little-endian it can reduce the cost, speed, code size, and power consumption and increasing performance. We have used two ASIP simulator like SimpleScalar and VEX. SimpleScalar simulator is an ASIP simulator; it consists of compiler, assembler, linker and simulation tools for the Simple Scalar PISA and Alpha AXP architectures. SimpleScalar tool set contains many simulators ranging from a fast functional simulator to a detailed out-of-order issue processor with a multi-level memory system. SimpleScalar also provides extensible, portable, high-performance architecture for high performance embedded systems design. Specific application compiled with using SimpleScalar, which generates application specific cache results. Another kind of ASIP simulator is VEX defines a parametric space of architecture that share a common set of application and system resources. VEX is a 32-bit clustered VLIW ISA is scalable and customizable to specific application domains.

## 2. RELATED WORK
Jain, M. K., Balakrishnan M. and Kumar A. proposed [1] scheduler based technique for exploring the register windows and cache configuration. Kin, J., Gupta, M. And Mangione-Smith, W. H. [2] analyzed energy efficiency by filtering cache references through an unusually small first level cache. A second level cache, similar in size and structure to a conventional first level cache, is positioned behind the filter cache and serves to mitigate the performance loss. Performance for different register file sizes is estimated by predicting the number of memory spills and its delay. Vivekanadarajah K. and Thambipillai, S. [3] the tuning filter cache to the needs of a particular application can save power and energy. Beside, a simple loop profiler directed methodology to deduce the optimal or near-optimal filter cache is proposed, without having to simulating all possible combinations of cache parameters from the specified space. The technique employed does not require explicit register assignment. Shuie, W. T. [4] Proposed three performance metrics, such as cache size, memory access time and energy consumption. Extensive experiments indicate that a small filter cache still can achieve a high hit rate and good performance. This approach allows the second level cache to be in a low power mode most of the time, thus resulting in power savings. Prikryl Z., Kroustck I., Hruska, T. and Kolar, D. [5] proposed automatically generated just-in-time translated simulator with the profiling capabilities. Gremzow, C. [8] using virtual machine architectures for ASIP synthesis and quantitative global data flow analysis for code partitioning, several "real world" applications from the domain of digital video signal processing. D. Fischer, J. Teich, M., Weper, R. [9] designed an efficient exploration algorithm for architecture/compiler co-designs of application-specific instruction-set processors. Guzman, V., Bhattacharyya, S.S, Kellomaki, E. and Takala, J. [10] developed an integration of SDF- and ASIP-oriented design flows, and use this integrated design flow to explore trade-offs in the space of hardware/software implementation and explore an approach to ASIP implementation in terms of "critical" and "non-critical" applications.

## 3. SIMPLESCALAR SIMULATOR
SimpleScalar simulator [6] used the MIPS architecture and support both big-endian and little-endian executable. SimpleScalar used the target files big-endian and little endian architecture is ssbig-na-sstrix and sslittle-na-sstrix, respectively. We have determined endian to our host environment and run the endian program located in the simplesim-2.0/ directory. SimpleScalar simulator provides fast cache simulation. SimpleScalar simulator is target specific simulator we have used 32-bit system as i-386 or 64-bit as i-686 host platform after targeting little-endian we have analyzed the cache memory result. In SimpleScalar we have used various application benchmarks and compiled with SimpleScalar version of GCC, which generates SimpleScalar assembly. The SimpleScalar assembly and loader, along with the necessary ported libraries, it produce SimpleScalar executable that can then be feel directly one of the provided simulators (this simulator compiled with the host's platforms) (see Figure 1).Simulator resources such as Sim-Cache,Sim-Safe etc. used for simulation.

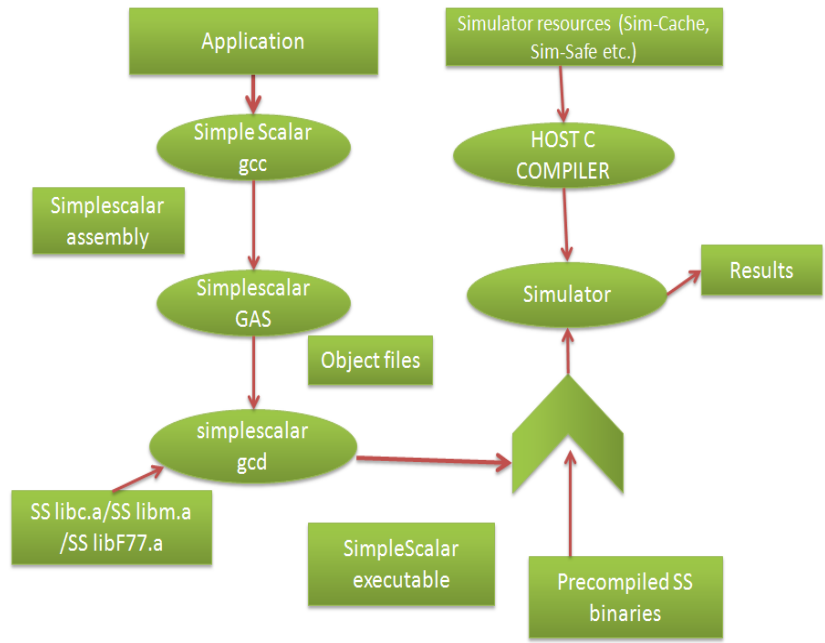

**Fig 1: SimpleScalar simulation overview**

## 3.1 SimpleScalar internals processor simulator

SimpleScalar simulator [6] contains five executions driven processor simulator. SimpleScalar processor simulator performs the non-blocking cache and speculative execution. We have used Sim-Cache for cache simulation.

Executions driven processor simulator are:

### 3.1.1 Sim-safe

This simulator is a functional simulation, it can providing alignment and access permissions for each memory reference. It contains the details of max instruction & scheduling operation. Complete simulation details show in Figure 2.

**Fig 2: Sim-safe simulator**

*3.1.2 Sim-cache:*
Simplescalar Sim-cache simulator performs the cache mapping as a set-associative mapping. Set associative mapping, is an improvement over the direct-mapping organization in that each word of cache can store two or more word of memory under the same index address. Each data word is stored to-gether with it's tag and no. of tag item in one word of cache is said to form a set. With the help of Sim-Cache we have implemented memory configuration according to specific application. This cache simulator performs the cache memory related results such as cache name, sets, cache associativity, cache block size, cache replacement policy etc. (see Figure 3).

Sim-cache used various cache configurations are:

-cache:dl1 <config> configures a level-one data cache.
-cache:dl2 <config> configures a level-two data cache.
-cache:il1 <config> configures a level-one instr. cache.
-cache:il2 <config> configures a level-two instr. cache.
-tlb:dtlb  <config>     configures the data TLB.
-tlb:itlb  <config>     configures the instruction TLB.
-flush     <boolean>    flush all caches on a system call;
-pcstat <stat>          generate a text-based profile.

**Fig 3: Sim-cache overview**

*3.1.3 Sim-cheetah*
Sim-Cheetah cache simulation engine to generating simulation results for multiple cache configurations with a single simulation. It's full associative efficiently as well as simulating a sometimes optimal replacement policy.

*3.1.4 Sim-profile*
SimpleScalar Sim-profile simulator generates detailed profiles on instruction classes and addresses, text symbols, memory accesses, branches, and data segment symbols.

*3.1.5 Sim-out order:*
SimpleScalar simulator supports the out-of-order processor's memory system which employs a load/store queue. Store values are placed in the queue and Loads are dispatched to the memory system when the addresses of all previous stores are known. Loads may be satisfied either by the memory system or by an earlier store value residing in the queue, if their addresses match. We can easily implementation with memory and processor by Sim-out order processor simulator.

*We can specify the processor core parameters are*

-fetch: ifqsize<size>   set the fetch width to be <size> instructions.
-Fetch: speed<ratio>
-fetch: mplat <cycles> set the branch misprediction latency.
-decode: width <insts> set the decode width to be <insts>, which must be a power of two.

*-issue: width <insts>    set the maximum issue width in a given cycle.*
*-issue: inorder        force the simulator to use in-order issue.*
*-issue: wrongpath      allow instructions to issue after a misspeculation.*
*-ruu:size <insts>      capacity of the RUU (in instructions).*
*-lsq:size<insts> capacity of the load/store queue (in instructions).*
*-res:ialu<num>         specify number of integer ALUs. -res:imult<num>     specify number of integer multipliers/dividers.*
*-res: memports<num> specify number of L1 cache ports.*

*-res:fpalu  <num>        specify number of floating point*
*-res: fpmult <num>       specify number of floating point*

*We can specify the memory hierarchy parameters are*

*-cache:dl1lat <cycles> specify the hit latency of the L1 data cache.*
*-cache:dl2lat <cycles>specify the hit latency of the L2 data cache.*
*-cache:il1lat <cycles> specify the hit latency of the L1 instruction cache.*
*-cache:il2lat <cycles> specify the hit latency of the L2 instruction cache.*
*-mem:lat <1st><next> specify main memory access latency (first, rest).*
*-mem: widths<bytes>     specify width of memory bus in bytes.*
*-tlb:lat<cycles>         specify latency (in cycles).*

## 3.2  SimpleScalar Cache Simulation

SimpleScalar simulator is an application specific Simulator which can produce the target specific cache memory results (see Figure 4). This SimpleScalar simulator contain cache simulator; this simulator can emulate a system with multiple levels of instruction and data caches. After simulation of SimpleScalar we can get the parameter such as total no. of instructions, sim-mem ref., sim-elapsed time, sim-inst-rate etc. (see Table [1]). We have analyzed the memory references according to the total no. of instruction executed (see Figure 5). Our model assumes two level data cache. Simplescalar tool suite defines both little-ness and big-endian-ness (target) of the architecture to improve the portability (the host machine is the one that matches the endian-ness of the host). A lot of features are available but we have used some limited parameter for ASIP simulation.

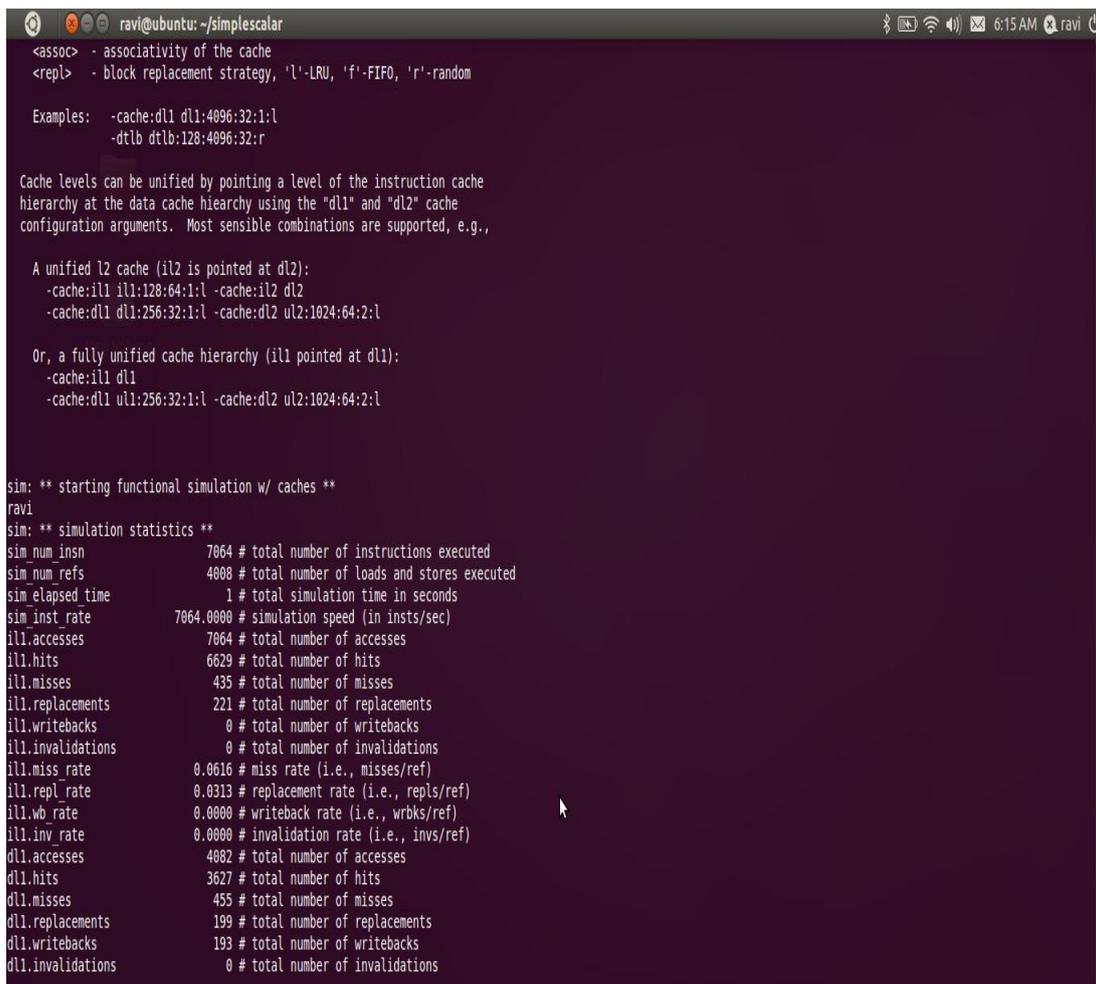

**Fig 4:  SimpleScalar Simulation Results**

**Table1. Simplescalar Simulation Result**

| Benchmarks | Total no. Of instruction | Sim-Memory-Ref. | Sim –elapsed time (sec) | Sim-inst-rate (inst/sec) |
|---|---|---|---|---|
| ll1.c | 5086757 | 1248274 | 3 | 1695585.6 |
| llM.c | 670697 | 162635 | 2 | 335348.5 |
| llMx.c | 7035 | 3936 | 1 | 7035.00 |
| llCs.c | 8339 | 4292 | 1 | 8339.0 |
| ll12.c | 8076 | 4263 | 1 | 8076.0 |

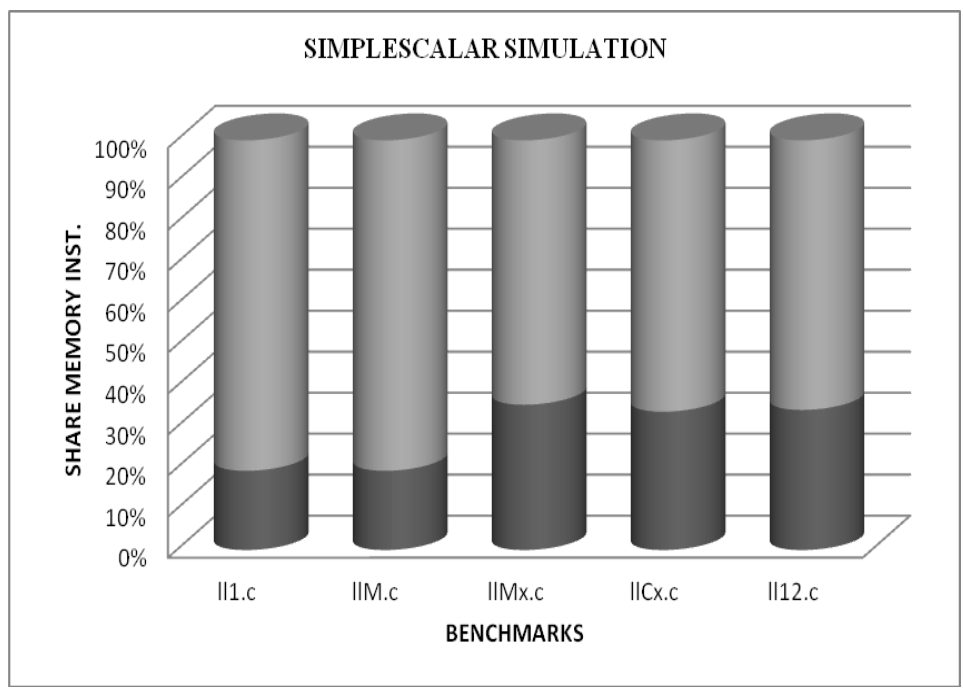

**Fig 5: Simplescalar simulation analysis with total no. of instruction and memory references**

## 3.3 Simplescalar limitation and features

Simplescalar tool set have following limitations.

- Simplescalar have no architectural delay slots : loads, stores and controll transfer do not executes the succeding instruction.
- Two level cache analysis.
- Target as little-endian and big-endian.
- Specific with host platform.

## 4. VEX SIMULATOR

VEX [7] provides a parametric space of architecture that share a common set of application and system resources, such as registers and operation.VEX is a 32-bit clustered VLIW ISA which is scalable and customizable to individual application. VEX simulator is an architecture-level (function) simulator that uses compiled simulator technology to achieve a speed of many equivalent 'MIPS'. This simulation system used sets of POSIX –like libc and libm libraries, VEX uses a cache simulator (level-1 cache only), and an API that enables for modeling the memory systems. VEX contains two qualifiers that specify streaming access (access to object that only exhibit spatial locality); and local access (access to object that exhibit a strong temporal locality).

### 4.1 VEX Cluster architecture

VEX uses cluster architecture (see Figure 6): it provides scalability of issue width and functionality using modular execution clusters. Each cluster is a collection of register files and a tightly coupled a set of functional units. Functional units within a cluster directly access only cluster register files. Data cache port and private memories are associated with each cluster. VEX allow multiple memory access to executes simultaneously.

### 4.2 Customization of VEX

VEX used load/store architecture, meaning that only load and store operations can access memory, and that memory operations only target general-purpose registers. VEX generally uses a big-endian byte ordering target model.

#### 4.2.1 VEX Cache customize

We can easily choose the cache configuration according to desire application. Vex contains various cache property such as cache size, sets, line size, no. of cache line size, cache miss penalty etc. (see Figure 7). In VEX compiled simulator we can easily specify the execution –driven parameters, such as clock and bus cycle, cache parameters (size, associativity, refill latency).

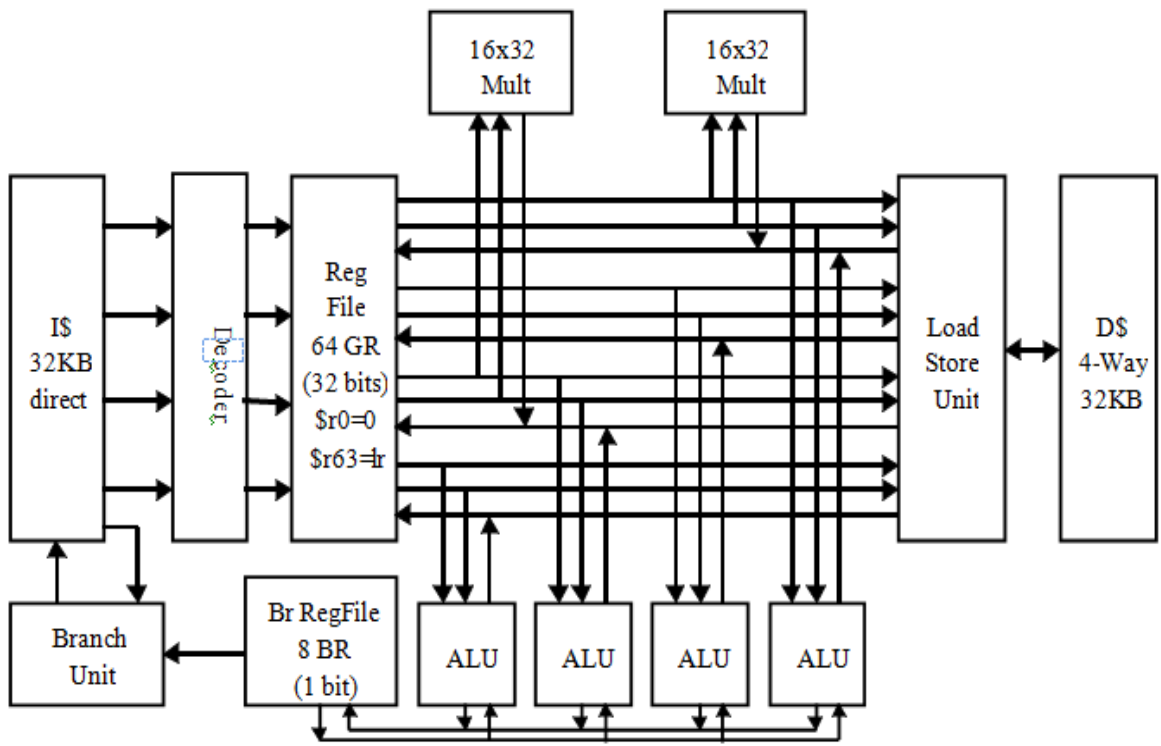

**Fig 6: VEX cluster structure**

```
vex.cfg
CoreCkFreq          1000
BusCkFreq           500
lg2CacheSize        16  # (CacheSize      = 256k)
lg2Sets             2   # (Sets           = 4)
lg2LineSize         5   # (LineSize       = 32)
MissPenalty         36
WBPenalty           33
lg2StrSize          9   # (StrSize        = 512)
lg2StrSets          4   # (StrSets        = 16)
lg2StrLineSize      5   # (StrLineSize    = 32)
StrMissPenalty      36
StrWBPenalty        33
lg2ICacheSize       15  # (ICacheSize     = 32k)
lg2ICacheSets       0   # (ICacheSets     = 1)
lg2ICacheLineSize   6   # (ICacheLineSize = 64)
ICachePenalty       45
NumCaches           1
BranchStall         1
StreamEnable        FALSE
PrefetchEnable      TRUE
LockEnable          FALSE
ProfGranularity     AUTO
```

**Fig 7: Customize Cache Parameter in vex.cfg file**

### 4.2.2 Vex cluster customize
The Defaults VEX cluster contains two register files, four integer ALUs, two 16x32-bit Multiply units, and a data cache port. The register set consists of 64 general purposes 32-bit registers (GRs) and 8 1bit branch register (BRs).

### 4.3 VEX VCG
VEX contain Visualization tools are often a very useful in the developemnt of various tunned application profiling and optimization of a complex application. This profiling usually necessary regardness of the target architecture. VEX have the rgg utility that converts the standards gprof output intoa VCG call graph (see Figure 8). Each Application can be implementing with VEX VCG and easily optimized specific application.

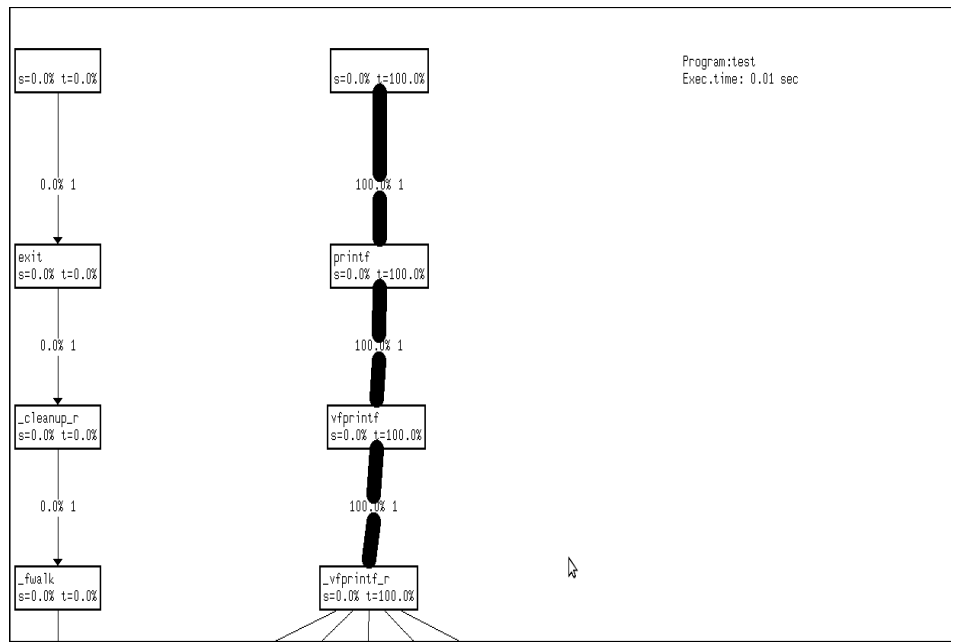

**Fig 8: VEX VCG**

## 4.4 VEX Cache Simulation

VEX development system (VEX tool chain) provides the set of tools that allow application benchmarks compiled for a VEX target to be simulated on a host workstation. VEX tool chain is mainly used for architecture exploration, application development, and benchmarking. It includes very fast architectural simulation that uses a form of binary translation to convert VEX assembler files. When we simulate an application with vex simulator it can generate assembly files (see Figure 9), Assembly files are simulated and we get execution statistics including cache misses. The pcntl utility is used for this purpose.

```
 Compress.s
.trace 1
L0?3:
        c0      add $r0.8 = $r0.5, 2    ## bblock 1, line 12-2,  t122, t232, 2(SI32)
        c0      add $r0.7 = $r0.2, 2    ## bblock 1, line 12-2,  t218, t179, 2(SI32)
        c0      add $r0.6 = $r0.5, 1    ## bblock 1, line 12-1,  t220, t232, 1(SI32)
        c0      add $r0.3 = $r0.2, 1    ## bblock 1, line 12-1,  t219, t179, 1(SI32)
;;                                      ## 0
        c0      sh2add $r0.3 = $r0.3, $r0.4   ## bblock 1, line 12-1,  t235, t219, t0
        c0      sh2add $r0.7 = $r0.7, $r0.4   ## bblock 1, line 12-2,  t236, t218, t0
        c0      add $r0.13 = $r0.5, 3   ## bblock 1, line 12-3,  t70,  t232, 3(SI32)
        c0      add $r0.12 = $r0.2, 3   ## bblock 1, line 12-3,  t94,  t179, 3(SI32)
;;                                      ## 1
        c0      sh2add $r0.12 = $r0.12, $r0.4 ## bblock 1, line 12-3,  t237, t94,  t0
        c0      add $r0.15 = $r0.5, 4   ## bblock 1, line 12-4,  t74,  t232, 4(SI32)
        c0      add $r0.14 = $r0.2, 4   ## bblock 1, line 12-4,  t212, t179, 4(SI32)
        c0      cmplt $b0.0 = $r0.5, 4  ## bblock 1, line 8-1,   t239(I1), t232, 4(SI32)
;;                                      ## 2
        c0      sh2add $r0.14 = $r0.14, $r0.4 ## bblock 1, line 12-4,  t238, t212, t0
        c0      add $r0.17 = $r0.10, 1  ## [spec] bblock 22, line 12-1, t128, t177, 1(SI32)
        c0      add $r0.16 = $r0.11, 1  ## [spec] bblock 22, line 12-1, t227, t178, 1(SI32)
        c0      add $r0.18 = $r0.11, 2  ## [spec] bblock 22, line 12-1, t225, t178, 2(SI32)
;;                                      ## 3
        c0      sh2add $r0.16 = $r0.16, $r0.4 ## [spec] bblock 22, line 12-1, t266, t227, t0
        c0      add $r0.19 = $r0.10, 2  ## [spec] bblock 22, line 12-1, t226, t177, 2(SI32)
        c0      sh2add $r0.18 = $r0.18, $r0.4 ## [spec] bblock 22, line 12-1, t267, t225, t0
        c0      add $r0.20 = $r0.11, 3  ## [spec] bblock 22, line 12-1, t223, t178, 3(SI32)
;;                                      ## 4
        c0      add $r0.21 = $r0.10, 3  ## [spec] bblock 22, line 12-1, t224, t177, 3(SI32)
        c0      sh2add $r0.20 = $r0.20, $r0.4 ## [spec] bblock 22, line 12-1, t268, t223, t0
        c0      add $r0.23 = $r0.10, 4  ## [spec] bblock 22, line 12-1, t137, t177, 4(SI32)
        c0      add $r0.22 = $r0.11, 4  ## [spec] bblock 22, line 12-1, t138, t178, 4(SI32)
;;                                      ## 5
        c0      sh2add $r0.22 = $r0.22, $r0.4 ## [spec] bblock 22, line 12-1, t269, t138, t0
        c0      mov $r0.24 = $r0.8      ## [spec] bblock 22, line 12-2, t85,  t122
```

**Fig 9: VEX assembly file**

```
Flat profile (cycles)
    Total  Total%    Insts   Insts%   Dcache  Dcache%   Icache  Icache% Function
     1958   22.37      396    26.61      396    13.52     1080    15.68 _vfprintf_r
      856    9.78       95     6.38       72     2.46      675     9.80 __sfvwrite
      578    6.60       56     3.76      108     3.69      360     5.23 __smakebuf
      457    5.22       64     4.30       72     2.46      315     4.57 _malloc_r
      434    4.96      238    15.99       36     1.23       90     1.31 mbtowc
      407    4.65      121     8.13        0     0.00      270     3.92 fflush
      402    4.59       36     2.42       -0    -0.00      360     5.23 _bcopy
      390    4.46       45     3.02      252     8.61       90     1.31 std
      318    3.63       49     3.29       36     1.23      225     3.27 _morecore_r
      297    3.39       25     1.68      180     6.15       90     1.31 printf
      277    3.17       47     3.16       -0    -0.00      225     3.27 __swsetup
      265    3.03       18     1.21      108     3.69      135     1.96 _fstat_r
      250    2.86       21     1.41        0     0.00      180     2.61 __sinit
      248    2.83       27     1.81       36     1.23      180     2.61 exit
      199    2.27       54     3.63       -0    -0.00      135     1.96 _fwalk
      189    2.16       49     3.29       -0    -0.00      135     1.96 __wrap_memchr
      186    2.13       48     3.23       36     1.23       90     1.31 _sbrk_r
      182    2.08        9     0.60       36     1.23       90     1.31 vfprintf
      161    1.84       22     1.48       -0    -0.00      135     1.96 __swrite
      159    1.82       20     1.34        0     0.00      135     1.96 _write_r
      136    1.55        8     0.54       36     1.23       90     1.31 localeconv
      106    1.21       13     0.87       -0    -0.00       90     1.31 __sprint
      104    1.19       11     0.74        0     0.00       90     1.31 __wrap_memmove
       54    0.62        7     0.47        0     0.00       45     0.65 _cleanup_r
       48    0.55        2     0.13        0     0.00       45     0.65 _localeconv_r
       90              36               45      (others not profiled)
```

**Fig 10: VEX D-CACHE and I-CACHE simulation analysis**

```
Total Cycles:                      8751 (0.017502 msec)
Execution Cycles:                  1488 ( 17.00%)
Stall Cycles:                      7263 ( 83.00%)
Nops:                               298 (  3.41%)
Executed operations:               1689

Executed branches:                  334 ( 19.78% ops)(22.45% insts)
Not taken branches:                  91 (  5.39% ops)(  6.12% insts)(27.25% br)
Taken branches:                     243 ( 14.39% ops)(16.33% insts)(72.75% br)
  Taken uncond branches:            147 (  8.70% ops)(  9.88% insts)(44.01% br)
  Taken cond branches:               96 (  5.68% ops)(  6.45% insts)(28.74% br)
Size of Loaded Code:              27264 Bytes

Instruction Memory Operations:
  Accesses:                        1250
    Hits (Hit Rate):               1130 ( 90.40%)
    Misses (Miss Rate):             120 (  9.60%)
Instruction Memory Stall Cycles
  Total (in cycles):               5580 (100.00%)
    Due to Misses:                 5400 ( 96.77%)
    Due to Bus Conflicts:           180 (  3.23%)

Data Memory Operations:           Cache
  Accesses:                         687 (100.00%)
    Hits (Hit Rate):                647 ( 94.18%)
    Misses (Miss Rate):              40 (  5.82%)
Data Memory Stall Cycles
  Total (in cycles):               1440 (100.00%)
    Due to Misses:                 1440 (100.00%)
    Due to Bus Conflicts:             0 (  0.00%)

Percentage Bus Bandwidth Consumed:  78.16%
```

**Fig 11: Vex Cache level1 simulation results**

VEX links with a simple cache simulation library, which models a L1 instruction and data cache memory. The cache simulator is really a trace simulator, which is embedded in the same binary for performance reasons. The VEX simulator supports for gprof, when invokes with the "-mas_G". Gprof running in the host environment. At the end of simulation, four files are created, *gmon.out* containing profile data that include cache simulation, *gmon-nocache.out* containing profile data not include cache simulation, *gmon-icache/gmon-dcache* containing data for respectively only instruction and data cache statistcs (see Figure 10).VEX gmon-icache/gmon-dcache file contains complete details of instruction and data memory operations,stall cycles,cache hit rate, cache miss rate etc.

VEX output file containing the complete statistcis, such as cycles (total,execution,stall,operations,time), branch statistics (execution, taken, condition, unconditions), instruction memory statistics (estimated codesize, hits/misses) data memory statistics (hits/misses, bus conflicts), bus statistics (bandwidth usages fration), simulation speed (mips,simulation time) (see Figure 11). In VEX cache simulation process we have used various standard benchmarks applications and after simulation we can get I-Cache & D-cache results according to the total no. of instruction executed ( see table(2,3)) and analyzed the I-cache/ D-cache according to desire application (see Figure 12).

**Table 2. VEX cache simulation results with total no. of instruction (I-CACHE/D-CACHE)**

| Benchmarks | D-cache values (mips) | I-cache values (mips) | total no. of instruction (mips) |
|---|---|---|---|
| Rgb_to_cmyk | 20.00 | 0.98 | 142.93 |
| Dither | 23.39 | 0.62 | 89.33 |
| Interpolate_x | 3.07 | 0.40 | 41.68 |
| Interpolate_y | 12.1662 | 0.231 | 34.70 |
| Ycc_rgb_converter | 1.818 | 0.057 | 24.57 |
| Jpeg_idct_islow | 0.51 | 0.02 | 5.54 |
| H2v2_fancy_upsample | 0.963 | 0.05 | 3.49 |
| Decode_mcu | 0.192 | 0.05 | 2.74 |
| Jpeg_fill_bit_buffer | 0.043 | 0.003 | 1.35 |
| Imppipe | 0.081 | 0.044 | 0.16 |

**Table 3. VEX cache simulation results (I-CACHE/D-CACHE)**

| Benchmarks | D-cache values (%) | I-cache values (%) |
|---|---|---|
| Rgb_to_cmyk | 13.9 | 0.6879664 |
| Dither | 26.1 | 0.6 |
| Interpolate_x | 7.3 | 0.9 |
| Interpolate_y | 35. | 0.6 |
| Ycc_rgb_converter | 7 | 0.2 |
| Jpeg_idct_islow | 9.1 | 0.3 |
| H2v2_fancy_upsample | 27.5 | 1.5 |
| Decode_mcu | 7.0 | 1.9 |
| Jpeg_fill_bit_buffer | 3.1 | 0.2 |
| Imppipe | 48.2 | 26.4 |

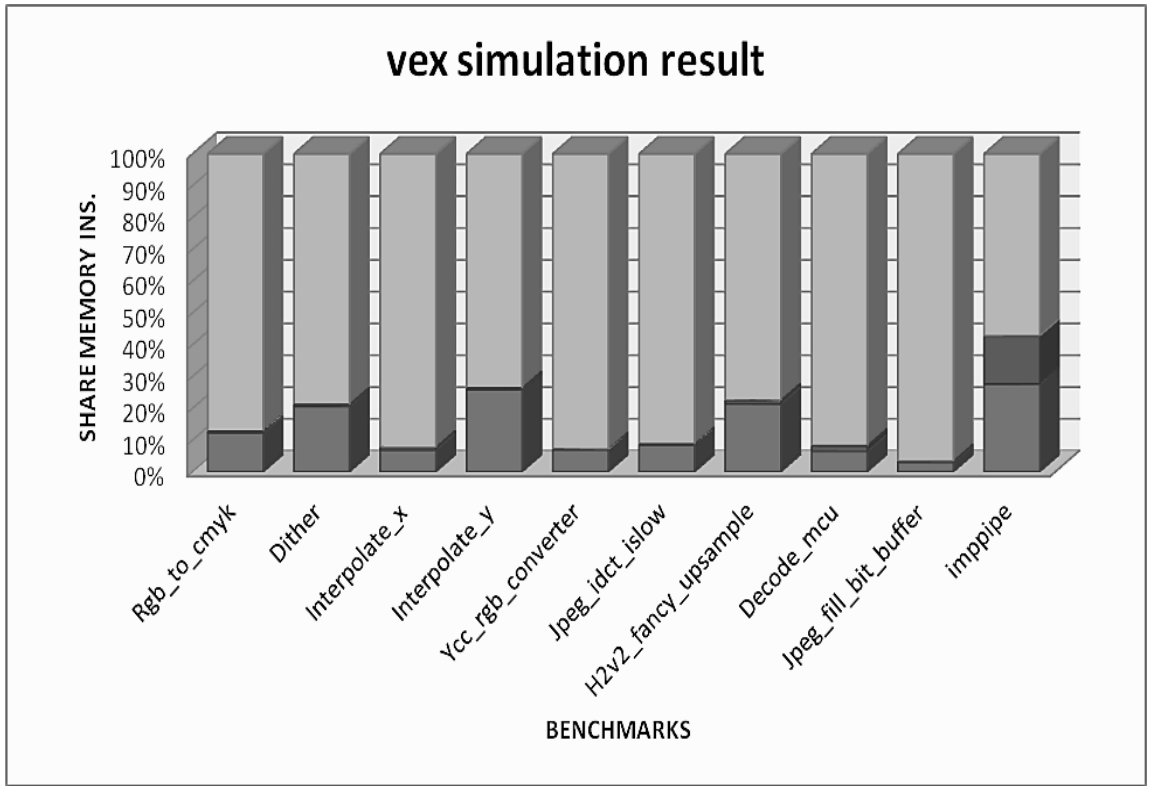

**Fig 12: Vex Cache Simulation result**

### 4.5 Vex features
Vex has following features.
- Cluster architecture.
- Single level cache analysis.
- Custom memory processor configuration.
- VLIW InstructionSetArchitecture.

### 5. CONCLUSION
ASIP simulators allows as the simulation of the cache memory design is an efficient manner. We have used ASIP simulators like SimpleScalar and VEX simulator performs target specific cache memory results. SimpleScalar is MIPS based architecture used in design space exploration, perform two level cache simulation.VEX defines a 32-bit clustered VLIW ISA is scalable and customizable to specific application and performs single level cache simulation. By the use of these ASIP simulators we have customized the memory configuration according to desire application and we can get complete details of I-Cache & D-cache according to the total no. of instructions executed.

### 6. ACKNOWLEDGMENTS
Our thanks to the SimpleScalar and VEX tool developer who has developed these simulators.